\documentclass[twocolumn,aps,pra,superscriptaddress,longbibliography]{revtex4-2}
\usepackage[utf8]{inputenc}
\usepackage[T1]{fontenc}
\usepackage[english]{babel}
\usepackage[dvipsnames]{xcolor}
\usepackage[colorlinks]{hyperref}
\usepackage[hypcap=true]{caption} 
\hypersetup{colorlinks,linkcolor=red,citecolor=blue,urlcolor=blue,final}
\usepackage{lmodern,fancyhdr}
\usepackage{graphicx,float}
\usepackage{enumitem,amsmath,amssymb,amsthm,bm,array,mathdots,mathrsfs,dsfont}
\usepackage{pifont}
\newcommand{\cmark}{\ding{51}}%
\newcommand{\xmark}{\ding{55}}
\newcolumntype{C}[1]{>{\centering\arraybackslash}m{#1}}

\newcommand{\colorlinks}[2]{{\hypersetup{allcolors=#1}#2}}

\newcommand{\be}{\begin{equation}}
\newcommand{\ee}{\end{equation}}

\newcommand{\tabref}[1]{Table~\ref{#1}}

\begin{document}

\title{Levitodynamics:
Levitation and control of microscopic objects in vacuum} 

\author{C. Gonzalez-Ballestero}
\affiliation{Institute for Quantum Optics and Quantum Information of the Austrian Academy of Sciences, A-6020 Innsbruck, Austria.}
\affiliation{Institute for Theoretical Physics, University of Innsbruck, A-6020 Innsbruck, Austria.}
\author{M. Aspelmeyer}
\affiliation{Vienna Center for Quantum Science and Technology, Faculty of Physics, University of Vienna, A-1090 Vienna, Austria}
\affiliation{Institute for Quantum Optics and Quantum Information, Austrian Academy of Sciences, A-1090 Vienna, Austria}
\author{L. Novotny}
\affiliation{Photonics Laboratory, ETH Zürich, 8093 Zürich, Switzerland}
\affiliation{Quantum Center, ETH Zürich, 8093 Zürich, Switzerland}
\author{R. Quidant}
\affiliation{Quantum Center, ETH Zürich, 8093 Zürich, Switzerland}
\affiliation{Nanophotonic Systems Laboratory, Department of Mechanical and Process Engineering,ETH Zurich, 8092 Zurich, Switzerland}
\author{O. Romero-Isart}
\affiliation{Institute for Quantum Optics and Quantum Information of the Austrian Academy of Sciences, A-6020 Innsbruck, Austria.}
\affiliation{Institute for Theoretical Physics, University of Innsbruck, A-6020 Innsbruck, Austria.}
\email{oriol.romero-isart@uibk.ac.at}

% \date{\today}

\begin{abstract}
    The control of levitated nano- and micro-objects in vacuum is of considerable interest, capitalizing on the scientific achievements in the fields of atomic physics, control theory and optomechanics. The ability to couple the motion of levitated systems to internal degrees of freedom, as well as to external forces and systems, provides opportunities for science and technology. Attractive research directions, ranging from fundamental quantum physics to commercial sensors, have been unlocked by the many recent experimental achievements, including motional ground-state cooling of an optically levitated nanoparticle. We review the status, challenges and prospects of levitodynamics, the mutidisciplinary research devoted to understanding, controlling, and using levitated nano- and micro-objects in vacuum.
\end{abstract}

\maketitle

\begin{figure*}[t!]
	\centering
	\includegraphics[width=0.7\linewidth]{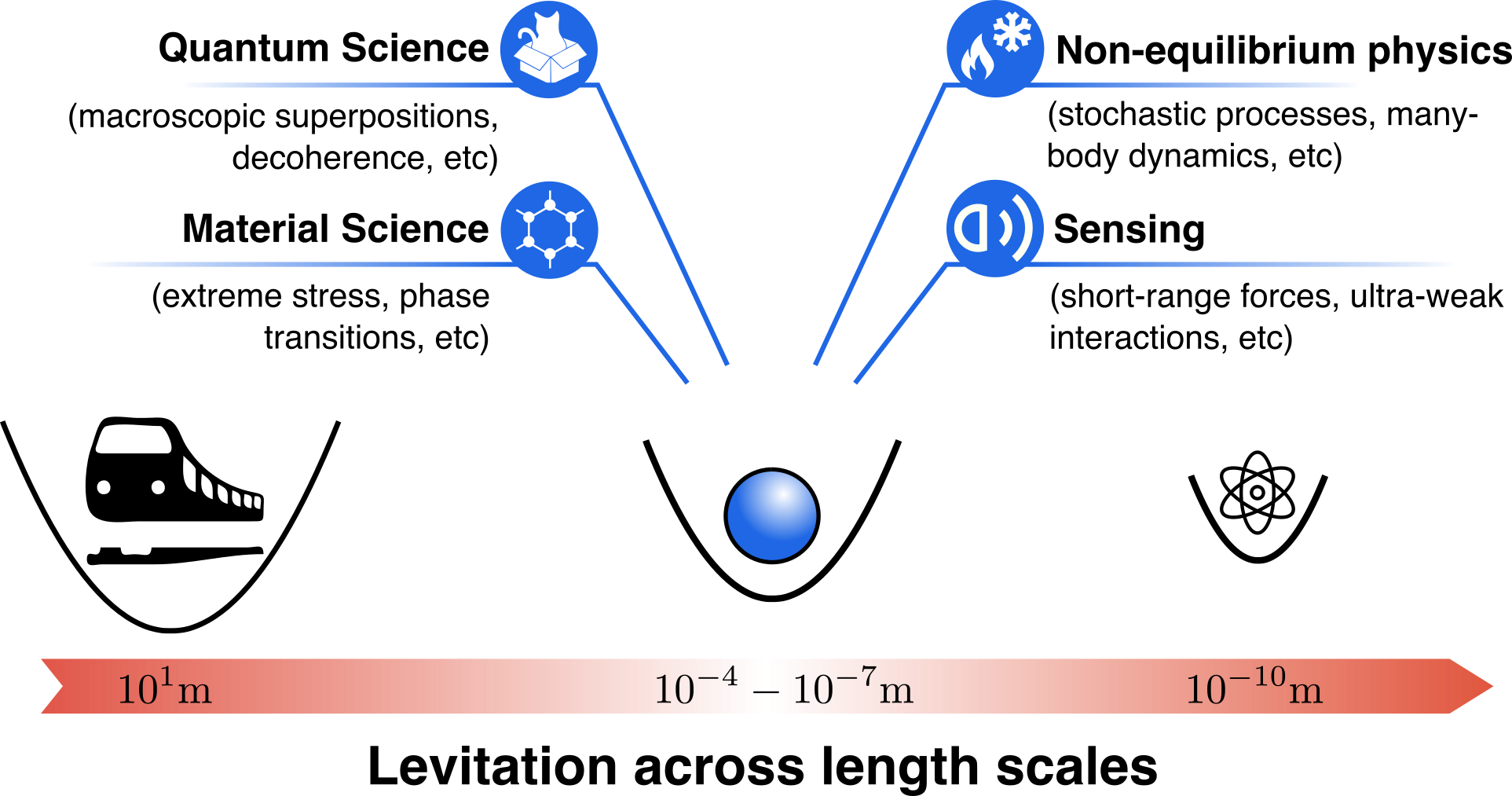}
		%\vspace{-0.2cm}
 	\caption{Controlled levitation of macroscopic- and atomic-scale objects has advanced technology and fundamental science, respectively. Levitodynamics—the levitation and control of microscopic solids in vacuum—targets objects in the intermediate-size regime. This research area provides access to unexplored research directions in quantum science, materials science, nonequilibrium physics, and sensing.}
\label{FigEnhancedAbstract}
\vspace{-0.25cm}
\end{figure*}

\section{Introduction}

\vspace{-0.25cm}

Levitation has been a topic of fascination for centuries. Although it was initially a subject reserved solely for science fiction, it became a reality by means of recent advances in science and technology~\cite{BrandtScience1989}. Well-known examples include magnetically levitated trains used for high-speed transportation and magnetically levitated centimeter-sized superconductors used as accelerometers~\cite{GoodkindRSI1999}. Further reducing the size of the levitated object enables a previously unattainable degree of control over its dynamics in vacuum. Indeed, in vacuum, the plethora of external (motion, rotation, libration) and internal (e.g., phonons, excitons, magnons, quantum impurities) degrees of freedom of a levitated micro- or nanoparticle can freely evolve or interact with each other in conditions of extreme environmental isolation. High-precision control over these degrees of freedom can be achieved by borrowing and building on techniques developed in atomic physics and biophysics~\cite{arnold1988spectroscopy,ChuPRL1986,FryeEPJQT2021,PhillipsRevModPhys1998,BustamanteNatRevMP2021,DholakiaNatPhot2011}, such as optical tweezers, dynamical control of trapping potentials, and motional cooling.

Levitodynamics—the trapping and control of nano- and micro-objects in vacuum—has established itself as an exciting research field by combining several ingredients from other fields in a single platform. When compared with other levitation platforms, two features stand out: (i) the large mass and density (and hence complexity) of the levitated object, as compared with the mass and density of trapped atoms; (ii) the high degree of control over both conservative dynamics and coupling to the environment, as compared with the degree of control for levitated meter-scale objects. The combination of these two features provides a springboard to many applications and branches of physics. For instance, large masses provide high inertial sensitivities for acceleration sensors. Moreover, the possibility of dynamically tuning both external potentials and dissipation provides a platform to study nonequilibrium physics and thermodynamics. This tunability also enables a top-down approach to quantum physics—namely, to bring a mesoscopic object of billions of atoms from the classical to the quantum regime to address long-standing questions about the nature of quantum mechanics at large scales. The possibilities of levitating particles of any material and of externally accessing their multitude of internal degrees of freedom also enable studies of condensed matter in unexplored regimes (e.g., in the absence of any substrate). Applications range from studying exotic material properties under extreme conditions to answering fundamental questions about the nature of equilibration in de facto isolated complex systems. Levitation also provides new opportunities for established fields—for instance, the study of liquid droplet levitation is useful for model development of aerosol processes in the atmosphere ~\cite{ReichCommPhys2020}. In essence, levitodynamics offers the possibility to levitate microscale objects of arbitrary shape and type in high vacuum and to study its motional and rotational dynamics, its internal properties, and the interplay between them.

In this Review, we detail advances in levitodynamics that have brought maturity to the field, as well as research directions currently under exploration. We discuss near-future prospects enabled by the untapped potential of levitodynamics and conclude with an overview of the main challenges that lie ahead.

\vspace{-0.3cm}

\section{Advances}
\vspace{-0.2cm}

Trapping of polarizable objects by means of optical forces was pioneered in the 1970s~\cite{AshkinPRL1970}. The subsequent development of optical tweezers revolutionized the field of biophysics~\cite{BustamanteNatRevMP2021}, whereas the levitation of microparticles~\cite{AshkinAPL1971,AshkinAPL1976} and, later, atoms in vacuum set the basis for modern atomic physics~\cite{ChuPRL1986,PhillipsRevModPhys1998}. Building on these advances, several theoretical works in the 2010s proposed applying quantum optical control techniques to levitated objects in vacuum~\cite{RomeroIsartNJP2010,ChangPNAS2010,BarkerPRA2010,RomeroIsartPRA2011b}. In this section, we review the advances sparked by the aforementioned works, which established levitodynamics as a vibrant research field.

\begin{figure*}[t!]
	\centering
	\includegraphics[width=\linewidth]{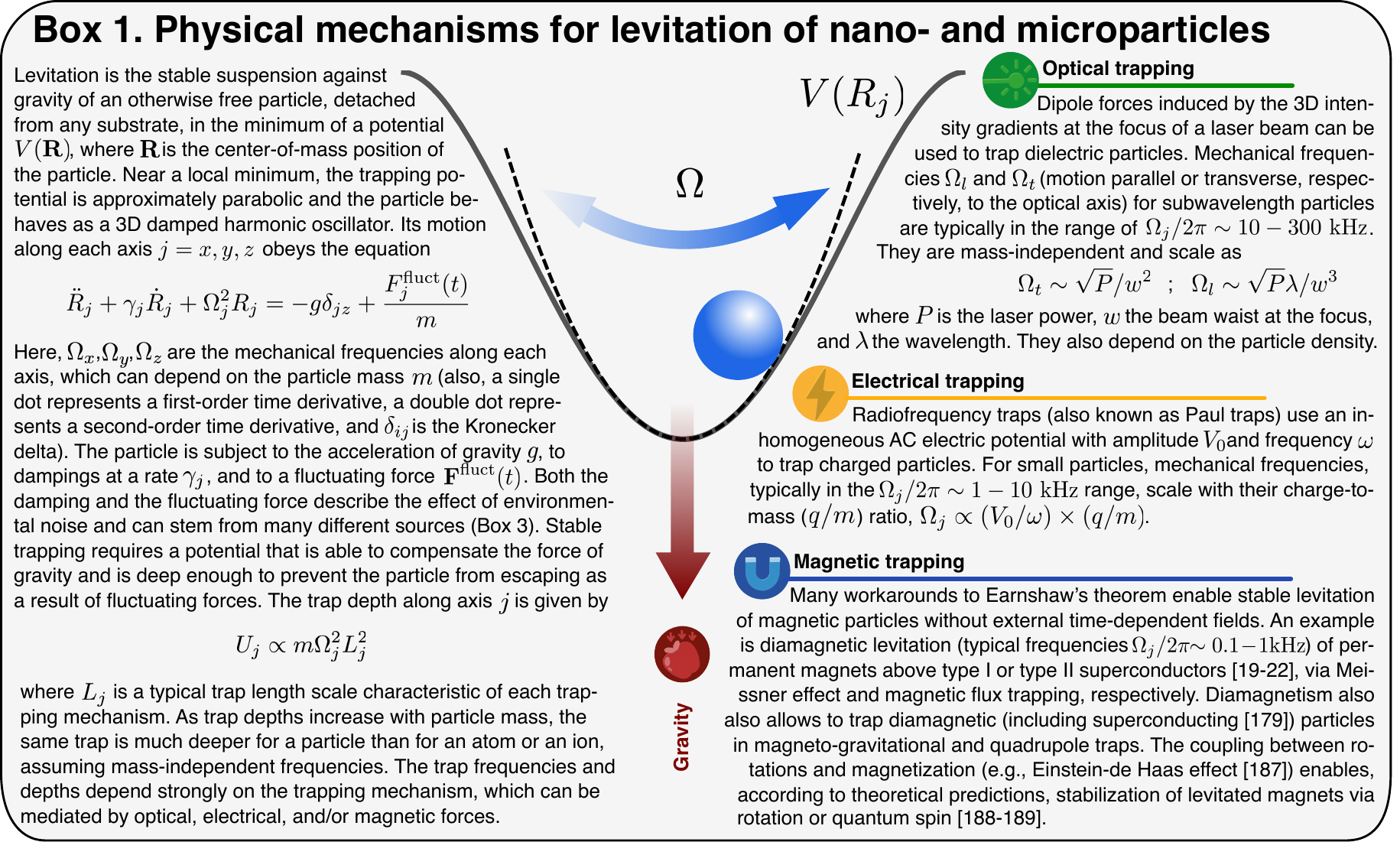}
	\vspace{-0.6cm}
	%\vspace{-0.2cm}
 	%\caption{\textcolor{purple}{CGB for myself: justify text, change particle color to blue, label ``Box 1: Levitation...''}}
 	\label{Box1Levitation}
\end{figure*}

\textit{Levitation schemes.} 
The physical mechanisms involved in the levitation of nano- and microparticles are summarized in Box ~\hyperref[Box1Levitation]{1}, and a comparison between different levitation methods is provided in~\tabref{Table}. Early levitation experiments made use of optical potentials and weakly absorbing dielectric polarizable particles. In the subsequent years, the toolbox expanded to include techniques borrowed from the atom-trapping community. The development of electrostatic and magnetic levitation made it possible to overcome excessive photoheating of the trapped specimen~\cite{MillenNatNano2014,RahmanSciRep2016} and extended levitation to magnets~\cite{GieselerPRL2020,VinantePhysRevAppl2020,TimberlakeAPL2019,WangPRAppl2019}, metals~\cite{SchellACSPhot2017}, diamond with quantum emitters~\cite{NeukirchNatPhot2015,HoangNatComm2016,ConanglaNanoLett2018,DelordPRL2018}, liquid droplets~\cite{HillPRL2008}, graphene flakes~\cite{NagornykhPRB2017}, and even superfluid helium droplets~\cite{BrownArxiv2021}.

\textit{Loading in high-vacuum.} 
Regardless of the trapping mechanism, one of the main challenges for trapping is the controlled loading of a particle into the trap in vacuum. Simple loading techniques such as spraying with a nebulizer rely on gas damping, which requires loading at higher pressures and subsequent pumping to desired vacuum conditions. More refined approaches enable particle loading in high vacuum. These methods make use of particle loading from a charged tip~\cite{DelordPRL2018} or particle launching with piezoelectric transducers~\cite{AshkinAPL1971}, electrospray~\cite{NagornykhAPL2015}, laser-induced acoustic desorption~\cite{KuhnNanoLett2015,BykovAPL2019,FlajvsmanovaSciRep2020}, or transfer through a load-lock~\cite{MestresAPL2015,CalamaiAIPAdv2021}.

\begin{figure*}[t!]
	\centering
	\includegraphics[width=\linewidth]{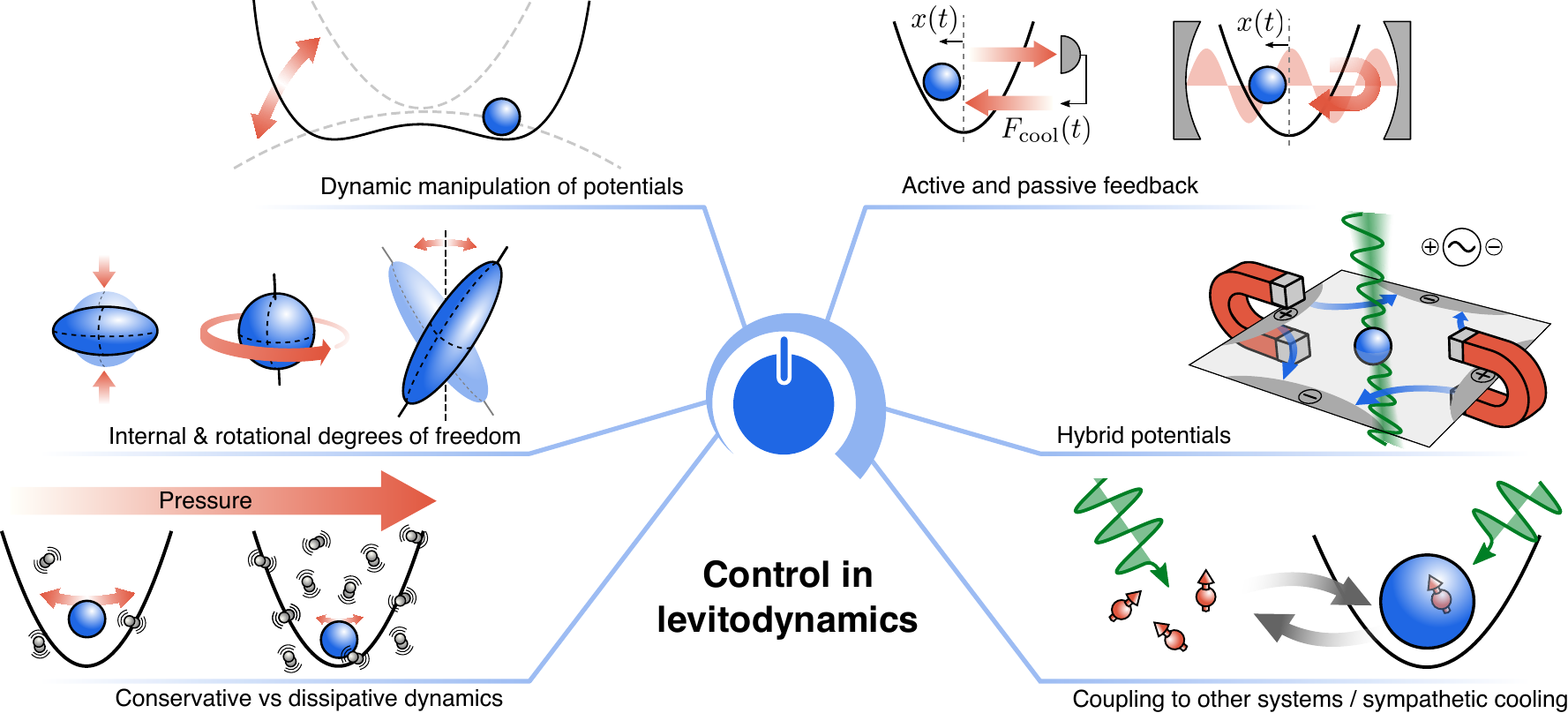}
	\vspace{0cm}
	%\vspace{-0.2cm}
 	\caption{Control in levitodynamics. The control toolbox of levitodynamics makes it possible to tailor every degree of freedom of a levitated particle in multiple ways. Both the conservative dynamics and the dissipation of motional, rotational, and \hspace{0.01cm} internal \hspace{0.01cm}(e.g., acoustic) degrees of freedom can be tailored via dynamic manipulation of potentials and \hspace{0.005cm} modification of \hspace{0.005cm} gas pressure, respectively. They can also be cooled \hspace{0.005cm} and parametrically driven via both active and \hspace{0.01cm}passive \hspace{0.01cm}feedback. \hspace{0.01cm}Additional \hspace{0.01cm} control \hspace{0.015cm}can \hspace{0.015cm}be \hspace{0.015cm}achieved \hspace{0.015cm}through \hspace{0.015cm}hybrid \hspace{0.015cm}traps \hspace{0.015cm}and \hspace{0.015cm}coupling \hspace{0.015cm}to \hspace{0.015cm}external \hspace{0.015cm}systems, which also provide controlled nonlinearity. $F_{\rm cool}(t)$, cooling force.}
\label{figcontrol}
\end{figure*}

\textit{Measurement and cooling.}
Another key advance has been the implementation of feedback control (cooling; see Fig.~\ref{figcontrol}) of levitated particle motion, which is essential for mitigating and controlling the impact of external noise. Freezing the motion of a levitated particle into the quantum ground state sets the stage for macroscopic quantum mechanics and has thus been a major objective in the field of levitodynamics. This quest has been pursued by means of both active- and passive-feedback cooling.

Active-feedback cooling relies on an actual measurement of the particle, the result of which is used to counteract particle motion by means of a control loop~\cite{LiNatPhys2011,GieselerPRL2012,HsuSciRep2016,SlezakNJP2018,GoldwaterQST2019,DaniaPRR2021}. The efficiency of active feedback relies on three factors: (i) quantum-limited measurement~\cite{JainPRL2016,KambaPRA2021}, (ii) optimal control~\cite{TebbenjohannsPRL2019,ConanglaPRL2019,IwasakiPRA2019}, and (iii) high detection efficiency~\cite{TebbenjohannsPRA2019}. The trade-off between measurement imprecision and measurement backaction is fundamentally constrained by the Heisenberg limit, which sets a lower bound on the performance of active-feedback cooling. Optimization on all three parameters has led to measurement near the Heisenberg limit, observation of motional sideband asymmetry~\cite{TebbenjohannsPRL2020}, and demonstration of quantum ground-state cooling for an optically levitated dielectric particle in both room temperature~\cite{MagriniNature2021} and cryogenic~\cite{TebbenjohannsNature2021} environments.

Passive-feedback cooling makes use of the autonomous feedback provided by an optical cavity~\cite{KieselPNAS2013,MillenPRL2015,MeyerPRL2019}. In the so-called resolved sideband limit, photons in the cavity most effectively counteract, and thereby slow, the motion of the particle by means of radiation pressure~\cite{RomeroIsartNJP2010,ChangPNAS2010,BarkerPRA2010,RomeroIsartPRA2011b}. Furthermore, as known from research into cavity cooling of atoms and molecules~\cite{HorakPRL1997,VuleticPRL2000}, laser noise can be minimized if the cavity is populated solely by the photons scattered from the particle (coherent scattering)~\cite{WindeyPRL2019,DelicPRL2019,GonzalezBallesteroPRA2019}. This configuration led to the first report on ground-state cooling of an optically levitated dielectric particle~\cite{DelicScience2020}.

\textit{Motional and rotational control.}
One of the main advantages of levitation is that the particles are free to move when presented with external influences, allowing for precise control of their external degrees of freedom (motion, rotation). Control of conservative particle motion is achieved via engineering of the trapping potential. On the one hand, complex static potentials can be synthesized using optical trapping beams with desired intensity profiles~\cite{RicciNatComm2017,AritaJOSAB2017,ZhouLasPhotRev2017,RondinNatNano2017} or by following a hybrid approach (i.e., combining optical and electric fields that act on charged dielectric particles)~\cite{MillenPRL2015,FonsecaPRL2016,ConanglaNanoLetters2020}. On the other hand, potentials can also be dynamically modified, either externally~\cite{NeukirchNatPhot2015,RashidPRL2016,HebestreitPRL2018,BykovAPL2019} -- including fast control of a complex optical potential~\cite{CiampiniArxiv2021} -- or by the backaction induced by the particle itself on the trapping field~\cite{KuhnOptica2017,NeumeierNJP2015}. Furthermore, the equilibrium position of a levitated particle can be controlled with high precision, enabling its positioning and translation with nanometer-scale accuracy. Such control has been implemented with moving single-beam traps~\cite{MestresAPL2015,DelicQST2020,CalamaiAIPAdv2021}, standing-wave traps~\cite{VcivzmarAPL2005,BrzobohatyNatPhot2013,GrassAPL2016}, and electric Paul traps~\cite{OstermayrRSI2018}.

%%%%%%%%%%%%%%%%%%%%%%%%%%

%%%%%%%%%%%%%%%%%%%%%%%%%%

%%%%%%%%%%%%%%%%%%%%%%%%%%

%%%%%%%%%%%%%%%%%%%%%%%%%%

%%%%%%%%%%%%%%%%%%%%%%%%%%

\begin{table*}[t]
\centering
%\captionsetup{justification=justified,singlelinecheck=false}
 \begin{tabular}{ C{2cm} || C{4.5cm} | C{4.5cm} | C{6cm} } 
   & Optical
   & 
   Electrical
   & 
   Magnetic
   \tabularnewline [2.5ex] 
 \hline\hline
 Levitation \textcolor{white}{Dummy text Dummy text Dummy text Dummy text Dummy text Dummy text Dummy text Dummy} & 
 \begin{enumerate}[wide, labelwidth=0.1pt, labelindent=4pt]
     \item[\large \textcolor{LimeGreen}{\cmark}] High trapping frequencies ($>100$kHz)
     \item[\large \textcolor{LimeGreen}{\cmark}] Charged or neutral particles
     \item[\large \textcolor{LimeGreen}{\cmark}/\large \textcolor{Red}{\xmark}] Large distance from surfaces
     \item[\large \textcolor{Red}{\xmark}] Laser recoil heating
     \item[\large \textcolor{Red}{\xmark}] Internal heating due to absorption
     \item[\large \textcolor{Red}{\xmark}] Limited to ultralow absorption, high-temperature-\allowbreak resistant particles
 \end{enumerate}
 \textcolor{white}{text Dummy textDummy text}
 & 
 \begin{enumerate}[wide, labelwidth=!, labelindent=4pt]
 \item[\large \textcolor{LimeGreen}{\cmark}] No internal heating 
 \item[\large \textcolor{LimeGreen}{\cmark}]
 Applicable to a wide range of charged particles
 \item[\large \textcolor{LimeGreen}{\cmark}]
 Large potential depth
 \item[\large \textcolor{LimeGreen}{\cmark}]
 On-chip integration
 \item[\large \textcolor{LimeGreen}{\cmark}]
 Compatible with optical manipulation
 \item[\large \textcolor{Red}{\xmark}]
 Low trapping frequencies \textcolor{white}{D} ($<10$kHz)
 \item[\large \textcolor{Red}{\xmark}]
 Highly charged particles (sensitive to decoherence)
 \item[\large \textcolor{Red}{\xmark}]
 Poor spatial confinement
 \end{enumerate}
 & 
\begin{enumerate}[wide, labelwidth=!, labelindent=4pt]
 \item[\large \textcolor{LimeGreen}{\cmark}] No internal heating
 \item[\large \textcolor{LimeGreen}{\cmark}]
 Potentials created by static fields
 \item[\large \textcolor{LimeGreen}{\cmark}]
 Applicable to very large particles
 \item[\large \textcolor{LimeGreen}{\cmark}]
  On-chip integration   
  \item[\large \textcolor{LimeGreen}{\cmark}/\large \textcolor{Red}{\xmark}]
  Advanced cryogenics required for superconducting traps or particles
  \item[\large \textcolor{Red}{\xmark}]
  Low trapping frequencies ($<1$ kHz)
  \item[\large \textcolor{Red}{\xmark}]
     Dissipation and decoherence mechanisms not fully understood
 \end{enumerate}
 \vspace{0.2cm}
 \textcolor{white}{Dummy text Dummy text Dummy Dummy text Dummy text Dummy text Dummy text Dummy text Dummy text}
  \tabularnewline  [-1ex] 
 \hline
 Control 
 \textcolor{white}{Dummy text Dummy text Dummy text Dummy text Dummy text}
 & 
 \begin{enumerate}[wide, labelwidth=!, labelindent=4pt]
 \item[\large \textcolor{LimeGreen}{\cmark}] Cooling via active and passive feedback
 \item[\large \textcolor{LimeGreen}{\cmark}]
 Potentials can be easily engineered and reconfigured
 \item[\large \textcolor{LimeGreen}{\cmark}]
 Control of rotational degrees of freedom
 \item[\large \textcolor{Red}{\xmark}]
 Coherent control limited by decoherence due to light scattering
 \end{enumerate} 
 & 
 \begin{enumerate}[wide, labelwidth=!, labelindent=4pt]
 \item[\large \textcolor{LimeGreen}{\cmark}] Cooling via active and passive feedback
 \item[\large \textcolor{LimeGreen}{\cmark}]
 Simple implementation of cold damping
 \item[\large \textcolor{LimeGreen}{\cmark}]
    Particles with internal degrees of freedom (quantum emitters)
\item[\large \textcolor{Red}{\xmark}]
     Proximity to surfaces (sensitive to stray fields)
 \end{enumerate}
 & 
 \begin{enumerate}[wide, labelwidth=!, labelindent=4pt]
 \item[\large \textcolor{LimeGreen}{\cmark}] Cooling via active and passive feedback
 \item[\large \textcolor{LimeGreen}{\cmark}]
Easy access to other degrees of freedom (e.g., magnetization)
\item[\large \textcolor{Red}{\xmark}]
     Static potentials
 \end{enumerate}
 \vspace{0.3cm}
 \textcolor{white}{Dummy text Dummy text Dummy Dummy text Dummy Dummy text DummyDummy text Dummy Dummy text Dummy}
 \tabularnewline [-1ex]  
 \hline
 Measurement 
 \textcolor{white}{Dummy text Dummy text Dummy }
 & 
\begin{enumerate}[wide, labelwidth=!, labelindent=4pt]
 \item[\large \textcolor{LimeGreen}{\cmark}] Shot-noise-limited measu-\allowbreak rements
 \item[\large \textcolor{Red}{\xmark}]
 Laser recoil heating
\end{enumerate} 
\vspace{0.5cm}
\textcolor{white}{Dummy text}
 & 
 \begin{enumerate}[wide, labelwidth=!, labelindent=4pt]
 \item[\large \textcolor{LimeGreen}{\cmark}] No need for bulk optical elements
 \item[\large \textcolor{Red}{\xmark}] Requires the detection of very weak currents
 \end{enumerate}
 \vspace{0.1cm}
 \textcolor{white}{Dummy text}
 & 
 \begin{enumerate}[wide, labelwidth=!, labelindent=4pt]
 \item[\large \textcolor{LimeGreen}{\cmark}] Sensitive readout with magnetometers (e.g., SQUIDs) or superconducting circuits (cavities, qubits)
  \item[\large \textcolor{Red}{\xmark}]
  Low compatibility with optical measurements   
 \end{enumerate} 
 \tabularnewline  [-1ex]  
 \hline
 \end{tabular}
 \caption{Advantages (\textcolor{LimeGreen}{\cmark}) and drawbacks (\textcolor{Red}{\xmark}) of optical, electrical, and magnetic interactions in terms of levitation, control, and measurement of the particle.}\label{Table}
 \vspace{-0.2cm}
\end{table*}

%%%%%%%%%%%%%%%%%%%%%%%%%%

%%%%%%%%%%%%%%%%%%%%%%%%%%

%%%%%%%%%%%%%%%%%%%%%%%%%%

%%%%%%%%%%%%%%%%%%%%%%%%%%

%%%%%%%%%%%%%%%%%%%%%%%%%%

Precise control has also been achieved over the rotational degrees of freedom of levitated particles. On the one hand, tuning the polarization direction of linearly polarized light on an optical trap allows external control of the equilibrium angular orientation of an anisotropic particle~\cite{KuhnOptica2017,PerdriatArxiv2021,SchellACSPhot2017}. On the other hand, using circularly polarized light, particles of different materials and shapes can be set into sustained rotation in a controlled fashion~\cite{NagornykhPRB2017,KuhnNanoLett2015,KuhnOptica2017,AritaNatComm2013,MonteiroPRA2018,KuhnNatComm2017,RashidPRL2018,HoangPRL2016}. Record spinning velocities in the gigahertz range have been reported for nanodumbbells~\cite{ReimannPRL2018,AhnPRL2018,AhnNatNano2020} and are limited only by particle disintegration when the elastic tensile limit is reached. Furthermore, rotational degrees of freedom can be cooled~\cite{DelordNature2020}, even simultaneously with the translational motion~\cite{BangPRR2020, VanDerLaanArxiv2020,SchaeferPRL2021}. Enhanced understanding of rotational noise~\cite{SticklerPRL2018}, especially noise induced by particle motion~\cite{VanDerLaanPRA2020}, has made it possible to reach the shot-noise limit, defined by radiation pressure torque~\cite{VanDerLaanArxiv2020}. This paves the way for bringing rotational degrees of freedom to the quantum regime.

\textit{Coupling to external systems.}
Experiments focused on coupling a levitated particle to external systems have been driven by the goals of (i) preventing photodamage in absorptive particles and (ii) exploiting nonlinear interactions (e.g., via coupling to a two-level system). Two prominent cases are levitated magnets coupled to other magnetic systems and diamond particles that contain color centers. Diamagnetically levitated magnets generate a position-dependent magnetic field that has been read out with both a superconducting quantum interference device (SQUID), which offers high sensitivity and minimal power dissipation at cryogenic temperatures~\cite{VinantePhysRevAppl2020,PratCampsPRAppl2017}, and a single electronic spin in a nitrogen-vacancy color center (NV center) embedded in a nearby diamond slab~\cite{GieselerPRL2020}. NV centers can also be embedded in a levitated nanodiamond~\cite{ConanglaNanoLett2018}, thereby equipping the particle with an internal energy-level structure. These quantum emitters can be optically manipulated and coherently controlled~\cite{HoangNatComm2016,DelordPRL2018}, even at the single-electron level~\cite{NeukirchNatPhot2015,SchellACSPhot2017}. Moreover, in the presence of a magnetic field gradient, electronic spins couple to the motion and rotation of the particle, as the Zeeman splitting of the electronic levels depends on the position and orientation of the particle. This spin-mechanical coupling in an ensemble of electronic spins has been used to control the orientation of a levitated diamond~\cite{PerdriatArxiv2021}, exert a torque~\cite{PelletArxiv2021}, and cool its librational motion via optical polarization~\cite{DelordNature2020}. Hosting quantum emitters in a levitated particle also provides a local probe for its internal dynamics.

\textit{Stochastic thermodynamics.}
The advances outlined above have enabled controlled levitation of micro- and nanosolids in high vacuum, advancing the use of levitodynamics in the pursuit of several research directions. The first example is the study of single-particle stochastic thermodynamics. Levitated particles provide a convenient platform for this goal, as motional and rotational degrees of freedom can be controlled, and their coupling to the environment can be tuned by many orders of magnitude. This tuning, a distinctive feature of levitation, can be performed by controlling gas pressure or by engineering artificial reservoirs (e.g., using an optical cavity)~\cite{DechantPRL2015}. The controllable isolation makes it possible to study equilibration~\cite{MillenNatNano2014}
or underdamped Brownian motion  both in the linear~\cite{LiScience2010} and nonlinear~\cite{GieselerPRL2014} regimes. As a specific example, the tunability of the damping rate facilitates exploration of thermally activated transitions between two potential minima separated by a barrier~\cite{RondinNatNano2017,RicciNatComm2017}. It has been shown that the transition probability is maximized at intermediate damping rates, a phenomenon predicted by Kramers in 1940. Dynamical modulation of trapping potentials is also a powerful tool. Specifically, it provides a means to study the equilibration from nonequilibrium initial states, enabling tests of fundamental fluctuation theorems~\cite{GieselerNatNano2014,HoangPRL2018}, and to implement shortcuts to adiabaticity for heat engine cycles~\cite{GieselerEntropy2018}. Along with feedback control, this technique also enables assessment of fluctuations under nonequilibrium thermal driving~\cite{RademacherArxiv2021} and of a generalized second law in non-Markovian information thermodynamics~\cite{DebiossacNatComm2020}. The combination of dynamical potential modulation and tunable dissipation provides a window into the thermodynamics of small systems, where fluctuations are relevant. The many open questions pertaining to this regime are explored, for example, by devising microscopic levitated heat engines~\cite{LiScience2010,GieselerEntropy2018} that can operate near the Curzon-Ahlborn efficiency bound~\cite{DechantPRL2015,DechantEPL2017}. In this realm, another quantum system, such as an optical cavity, enables gradual substitution of the thermal reservoir to experimentally study the extension of the concepts of classical stochastic thermodynamics into the quantum regime~\cite{AgarwalPRE2013}.

\textit{Interrogation of materials.}
Another feature of levitated nano- and microparticles is the ability to control and investigate mesoscopic solids composed of billions of atoms in conditions of extreme isolation. Important applications are measurement and reduction of a particle’s internal temperature. This is relevant for preventing particle loss at low pressures, reducing blackbody decoherence for fundamental tests of quantum mechanics, or reducing decoherence of embedded color centers for use in quantum transduction. One way to measure the internal temperature of optically levitated particles is via the mechanical frequency shift induced by a temperature-dependent refractive index~\cite{HebestreitPRA2018}. Another way is to use the fluorescence of embedded quantum emitters~\cite{HoangNatComm2016,DelordAPL2017}, which also allows for internal refrigeration of the particle~\cite{RahmanNatPhot2017}. Additionally, surface temperature can be estimated by tracking particle motion in the background gas~\cite{MillenNatNano2014}.

\begin{figure*}[t!]
	\centering
	\includegraphics[width=\linewidth]{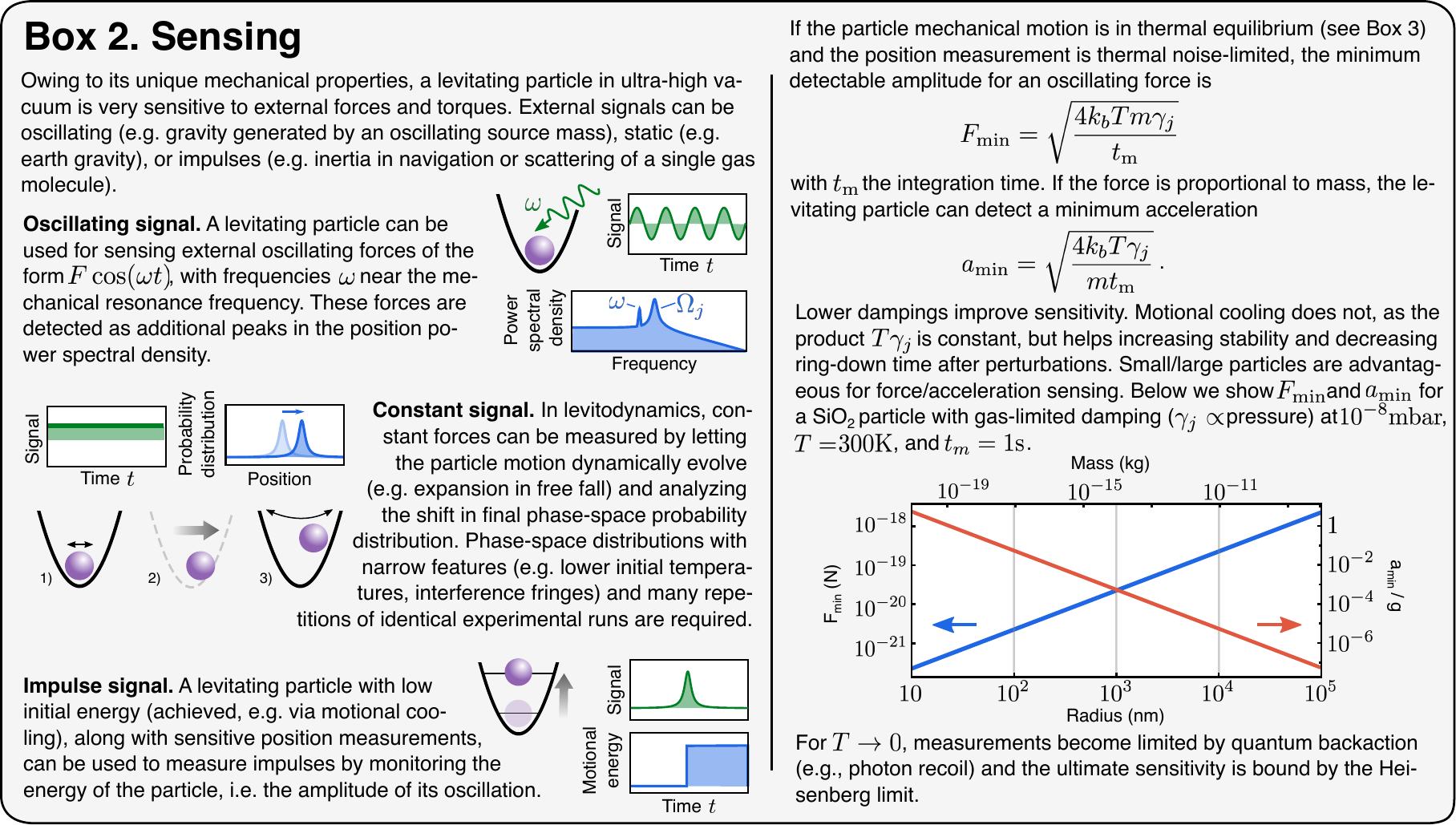}
	\vspace{-0.6cm}
	%\vspace{-0.2cm}
 	%\caption{\textcolor{purple}{CGB for myself: justify text, change particle color to blue, label ``Box 1: Levitation...''}}
\label{Box2Sensing}
\end{figure*}

A related research direction is the study of matter in controlled conditions (i.e., without the perturbing effects of any substrate). The absence of clamping allows particles to be rotated at frequencies that exceed 1 GHz~\cite{ReimannPRL2018,AhnPRL2018,MonteiroPRA2018}, which, for particles of $100$-nm radius, results in surface velocities of $600$ m/s and centrifugal accelerations of $4 \times 10^{11}g$ (where g is Earth’s gravitational acceleration). The extreme surface forces and material stresses provide information about material susceptibilities, their modification as a result of large centrifugal stress~\cite{NagornykhPRB2017}, and tensile failure~\cite{SchuckSciAdv2018}. The absence of a substrate has received particular attention in the study of liquid particle (droplet) physics, which is important for modeling a wide range of systems, from atomic nuclei to black holes~\cite{HillPRL2008}. Moreover, at low temperatures, levitation of liquid helium droplets provides insight into superfluid physics~\cite{ChildressPRA2017} -- e.g., into the complex interplay between fast droplet rotation and the formation of superfluid vortices. In this scenario, by providing controlled and stable trapping, levitation offers a definite advantage over current methods based on x-ray diffraction of liquid helium jets~\cite{LangbehnPRL2018}.

\textit{Sensing.}
Levitated particles also excel as force sensors (Box~\hyperref[Box2Sensing]{2}). Their three-dimensional (3D) motion and rotation allows for independent sensing of forces and torques along three dimensions (6D sensing) with shot-noise-limited sensitivity. Moreover, their large motional amplitude provides a much larger dynamic range than that afforded by clamped resonators. In addition, the spectral response of levitated sensors can be largely tuned by potential engineering, thus modifying their oscillation frequencies. Force sensitivities in the sub-femtonewton regime have been demonstrated~\cite{MoorePRL2014,RanjitPRA2016,HempstonAPL2017,LiNatPhys2011,GieselerNatPhys2013,GieselerPRL2012,RiderPRL2016}. The high force sensitivity can be exploited for the sensing of different fields. For example, if the levitated particle carries a charge, it responds to electric fields and thus constitutes an electric field sensor~\cite{MoorePRL2014,FrimmerPRA2017,RanjitPRA2016,SlezakNJP2018,RicciNanoLett2019}. Likewise, if the particle carries a magnetic moment it responds to magnetic fields and acts as a magnetic field sensor~\cite{KimballPRL2016,HoangPRL2016}. Furthermore, the high mass of a levitated particle (as opposed to, e.g., a trapped atom) makes it a natural sensor for gravitational and inertial forces, including acceleration and rotation (gyroscopes)~\cite{MonteiroPRA2017,HebestreitPRL2018,RiderPRA2018,TimberlakeAPL2019,MonteiroPRA2020}. Anisotropic particles, such as rods, dumbbells, or other particles composed of anisotropic materials, can be set into rotation and used for torque sensing with sensitivities exceeding $10^{-27}$N$\cdot$m/Hz${}^{-1/2}$~\cite{AhnNatNano2020,HoangPRL2016,RashidPRL2018,AhnPRL2018,VanDerLaanPRA2020}. 
In addition, a polarizable particle also responds to optical forces, such as radiation pressure, which can be measured with photon-recoil-limited precision~\cite{JainPRL2016}. Finally, the lack of a clamping mechanism allows levitated particles to be released from their trap and recaptured at a later time and different location. This feature opens up the possibility of measuring forces ``in the dark'' (e.g., the gravitational force in a free-fall experiment~\cite{HebestreitPRL2018}). Aside from sensing, this protocol is of interest for quantum matter-wave interferometry with massive, solid-state particles~\cite{RomeroIsartPRL2011,RomeroIsartPRA2011,KaltenbaekExpAstronom2012,BatemanNatCom2014,GeraciPRD2015,WanPRL2016,PinoQST2018}.

\section{Future research directions}

The potential of levitodynamics is largely untapped. However, enabled by the aforementioned advances, a large variety of research directions and applications will become attainable in the near future. We classify these directions in three broad, interconnected areas: (i) sensing and metrology, from commercial navigation sensors to searches for new physics such as dark matter; (ii) novel quantum physics, e.g., generation of macroscopic quantum superpositions; and (iii) physics of complex systems, from nonequilibrium physics to materials engineering.

\textit{Integrated sensors.}
On the applied side, levitation-based sensing may find uses in areas ranging from seismometry to inertial navigation systems. Development of commercial sensors, however, will require integration of trapping, control, and readout on a single platform to reduce noise and achieve long-term stability. The prospect of compact and robust integrated sensors motivates the ongoing effort toward on-chip levitodynamics. Controlled levitation near structured surfaces, a key first step, has already been achieved in ambient conditions and low vacuum. Dielectric nanoparticles have been levitated in planar Paul traps~\cite{AldaApplPhysLett2016}, near dielectric photonic structures~\cite{MagriniOptica2018} and metasurfaces~\cite{ShenArxiv2021}, and near ultrathin membranes~\cite{DiehlPRA2018}. Levitation of magnetic particles near the surface of superconductors~\cite{PratCampsPRAppl2017}, which has applications in high-sensitivity magnetometry~\cite{KimballPRL2016}, has also been reported~\cite{GieselerPRL2020,VinantePhysRevAppl2020,TimberlakeAPL2019,WangPRAppl2019}. Current efforts focus on reducing particle-surface separations, increasing vacuum conditions, and using the same nearby surfaces or structures for particle control and readout~\cite{KuhnApplPhysLett2017,WinstonePRA2018}. On this front, particles have already been levitated at subwavelength distances from surfaces~\cite{MagriniOptica2018,MontoyaArxiv2021}, even in high vacuum with the aid of feedback cooling~\cite{DiehlPRA2018}. Moreover, exploitation of the strong mode confinement of a photonic crystal cavity has enabled enhanced readout of the particle dynamics~\cite{MagriniOptica2018}. Controlling and mitigating the still poorly understood surface-induced noise, also known as anomalous heating in the trapped-ion community~\cite{BrownuttRMP2015}, remains a major challenge for on-chip levitodynamics.

\textit{Rotation-based sensing and metrology.}
Owing to their complex dynamics, the rotational degrees of freedom of a levitated particle can be exploited in many areas of sensing and metrology. Fast-rotating nanorotors can act as local magnetic fields~\cite{KimballPRL2016} or pressure sensors~\cite{KuhnNatComm2017}. Moreover, the frequency of optically levitated nanorotors can be locked to external clocks and tuned across a wide frequency range. The frequency stability of these nanorotors~\cite{VanDerLaanPRA2020} is limited only by clock stability, even at room temperature~\cite{KuhnNatComm2017}. This will enable their use as frequency standards for metrology. Finally, rotation can be used for sensing fundamental forces. Fast-rotating nanorotors have been proposed as torsion balances for measuring processes such as Casimir torques or vacuum friction~\cite{AhnPRL2018,AhnNatNano2020,ManjavacasPRA2010,ZhaoPRL2012}. Measuring such electromagnetic dispersion forces could elucidate the mechanisms behind surface-induced heating, opening the door for compact levitation platforms. Furthermore, controlled rotation will facilitate cancellation of limiting noise in fundamental sensing experiments, such as the measurement of Newton’s gravitational constant~\cite{WestphalNature2021} or the search for exotic particles beyond the standard model~\cite{MoorePRL2014}. A major challenge for rotational sensing and metrology is to reach the fundamental sensitivity limit imposed by the quantization of angular momentum and associated radiation torque shot noise~\cite{VanDerLaanArxiv2020}.

\begin{figure*}[t!]
	\centering
	\includegraphics[width=\linewidth]{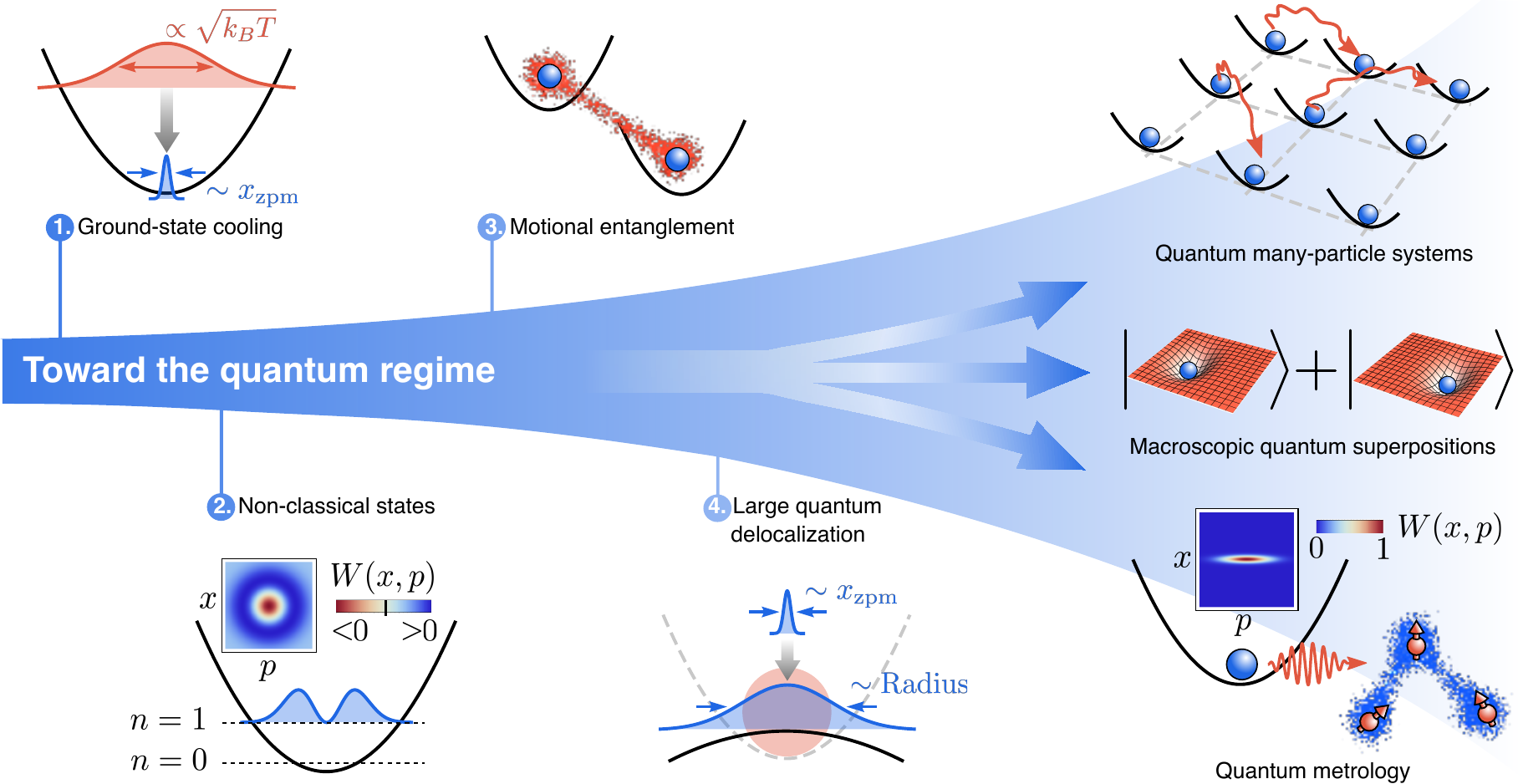}
	\vspace{0cm}
	%\vspace{-0.2cm}
 	\caption{Quantum levitodynamics. Levitated particles in the quantum regime open new research directions in fun-\allowbreak damental quantum mechanics, quantum sensing, and many-particle physics, among other areas. Reaching this regi-\allowbreak me is a challenging effort that must be undertaken sequentially. Some milestones along the way are (1) ground-state cooling, which reduces the thermal motional amplitude to the zero-point motion $x_{zpm}$ ($k_B$, Boltzmann constant); \hspace{-0.1cm}(2) generation of nonclassical motional states at length scales $\sim x_{zpm}$ (e.g., Fock states); (3) demonstration of entangle-\allowbreak ment between the motional degrees of freedom of two levitated particles; and (4) delocalization of the particle wave function to an extension orders of magnitude larger than the zero-point motion, comparable to its size.}
\label{FigureQuantum}
\end{figure*}

\textit{New physics.}
The possibility of bringing particles to the quantum regime provides a novel sensing platform for fundamental science. Specifically, the substantial degree of isolation and control required to reach the quantum regime yields laboratory-scale experiments that are extremely sensitive to forces or phenomena that couple to mass. A first possible use is the search for new physics beyond the standard model of particle physics. As an alternative approach to colliders pushing the energy frontier, these compact levitodynamics experiments push the sensitivity frontier~\cite{MooreQST2021}. They can exploit (i) the high isolation and mass density of levitated particles and (ii) levitation near structured surfaces~\cite{DiehlPRA2018} to detect tiny deviations arising from high-energy physics at short distances~\cite{GeraciPRL2010}. This will enable tests of short-distance corrections to Coulomb’s law to exclude dark matter models in the hidden sector, as well as searches for fundamental non-neutrality of matter~\cite{AfekPRD2021}. Experiments along this route have already been performed using optically levitated microparticles~\cite{MoorePRL2014}. Moreover, these experiments will allow researchers to test potential corrections to Newton's gravitational law at the unexplored micrometer and submicrometer scale~\cite{GeraciPRL2010,KawasakiRevSciInstrum2020,BlakemoreArxiv2021}, an order of magnitude beyond the 50-$\mu$m scale where strong constraints to deviations have been reported~\cite{LeePRL2020}. A second related prospect for these sensitive experiments is the detection of astrophysical signals, including high-frequency gravitational waves~\cite{AggarwalArXiv2020} and signatures of dark matter and dark energy~\cite{MoorePRL2014,BatemanSciRep2015,RiderPRL2016,MonteiroPRL2020,CarneyQSciRep2021}. Finally, levitated particles could shed light on the quantum nature of gravity by entangling two levitated massive objects via their mutual gravitational interaction~\cite{MarlettoPRL2017,BosePRL2017}. As discussed below, exploring all of these research directions requires devising advanced techniques to suppress quantum decoherence.

\textit{Quantum sensing and transduction.} Bringing a levitated particle to the quantum regime also enables a variety of genuine quantum applications (Fig.~\ref{FigureQuantum}), such as quantum-enhanced sensing. These applications require cooling of the particle to the quantum regime, as well as the ability to coherently couple its motion to other quantum systems—that is, to establish a coherent coupling at a rate higher than the decoherence rate of both the particle and the external quantum system. Although such strong coupling has been demonstrated with dielectric particles coupled to an optical cavity~\cite{DeLosRiosNatComm2021}, establishing coherent coupling to other quantum systems remains to be achieved. Two-level systems are particularly relevant for current research, owing to their large intrinsic nonlinearity~\cite{MartinetzNPJQI2020}. As in clamped quantum nanomechanical systems, coherent coupling to such nonlinear systems will enable the preparation of levitated particles in non-Gaussian quantum states -- that is, states characterized by a phase-space quasi-probability density with negative values. These states can be used to unequivocally certify the quantum nature of the particle’s motional state and to identify decoherence sources, to which they are extremely sensitive. Moreover, coherent coupling to an ensemble of entangled two-level quantum systems will improve the displacement readout of levitated particles, with potential uses in inertial or -- in the case of levitated magnets -- magnetic quantum sensing. More specific to levitodynamics are applications in which an interface with internal solid-state quantum excitations, such as acoustic phonons, could be established. Theoretical proposals~\cite{GonzalezBallesteroPRL2020} indicate that, owing to their nearly perfect isolation, these long-lived quantum excitations could be used, for instance, as quantum memories or quantum transducers.

\textit{Macroscopic quantum superpositions.} 
One of the most appealing features of quantum physics is the ability to prepare macroscopic superpositions, in which an object behaves as if it were in two locations at once, separated by a distance comparable to or larger than its size. Experimental confirmations started as early as 1927, using electrons, and have today reached the size of organic molecules that contain thousands of atoms~\cite{FeinNaturePhysics2019}. Levitodynamics will enable the preparation of macroscopic quantum superpositions of nanoparticles that contain billions of atoms and, in principle, far beyond. These superpositions will enable quantum matter-wave interferometry with mesoscopic particles~\cite{RomeroIsartPRL2011,RomeroIsartPRA2011,KaltenbaekExpAstronom2012,BatemanNatCom2014,SticklerNJP2018} and will corroborate the quantum superposition principle in a hitherto inaccessible parameter regime where collapse models predict the breakdown of quantum mechanics~\cite{RomeroIsartPRA2011,BassiRMP2013}. Moreover, they will serve as ultraprecise sensors able to detect tiny signals~\cite{SticklerNJP2018,GeraciPRD2015} and will thus promote enhanced understanding and testing of theoretical models of environmental decoherence. This is made possible by the extreme sensitivity of these superpositions to decoherence. Finally, superpositions will open the door for exploring the interplay between quantum mechanics and gravity -- for example, by elucidating the gravitational field generated by a massive object in a macroscopic quantum superposition state~\cite{Cecile2011ChapelConference}.

Macroscopic quantum superpositions require sufficiently pure states that are coherently delocalized over large scales. These conditions could be achieved by ground-state motional cooling, followed by large-scale expansion of the wave function through dynamical control of the trapping potential~\cite{PinoQST2018,RomeroIsartNJP2017,WeissPRL2021} (Fig.~\ref{FigureQuantum}). This expansion increases the spatial extent of the wave function by several orders of magnitude, from the zero-point motion in the ground state (typically around $10^{-12}$m) to, ideally, a size comparable to that of the particle itself ($10^{-7}$m). Several theoretical works have proposed and analyzed these ideas and have determined the requirements to prevent the detrimental action of decoherence from the environment~\cite{RomeroIsartPRL2011,RomeroIsartPRA2011,KaltenbaekExpAstronom2012,BatemanNatCom2014,WanPRL2016}. After the achievement of motional ground-state cooling~\cite{DelicScience2020,MagriniNature2021,TebbenjohannsNature2021}, the next steps consist of reaching the low decoherence level required for unitary expansion of the wave function and devising detection schemes to certify the successful preparation of a macroscopic superposition~\cite{WeissPRR2019}.

\textit{Non-equilibrium physics.}
Levitated particles offer a distinctive channel to probe and engineer complex systems across multiple scales. These encompass arrays of levitated particles for many-body physics, single particles for motional quantum heat engines, and solid-state physics inside a single particle. At the largest scale, ensembles of levitated particles can be used to explore many-body physics as done with atomic ensembles, with the addition of rotational degrees of freedom, rich internal solid-state structure, tunable coupling to the environment, and gravitational interactions, among others. On the one hand, ensembles of nanoparticles levitated in a single wide and deep trap will enable the exploration of nonequilibrium self-assembly, liquid crystal phases, and the transitions between them~\cite{DholakiaRMP2010,LechnerPRL2013}. Conversely, arrays of physically separated particles with tunable interactions can be tailored by multiplexing several traps into an array~\cite{BurnhamOptExp2006}. In these systems, the ability to tune the dissipation enables exploration of the transition between closed and open systems. This may contribute to previous studies~\cite{GringScience2012} of elusive phenomena, such as prethermalization~\cite{LiuAQT2020}, and tests of long-standing hypotheses, such as the eigenstate thermalization hypothesis.

At the single-particle level, levitodynamics has turned out to be an excellent approach to implement far-from-equilibrium processes and understand irreversibility, similar to the complementary and long-established approach of colloids in liquids. This will certainly not be limited to optical approaches in the future. With the first demonstrations of levitated systems in the quantum regime, new ground can be explored at the interface of classical, information, and quantum thermodynamics. Focusing on the refinement of levitated microheat engines as a paradigmatic example, the rapid control enabled (e.g., by all-optical approaches) theoretically designed optimal protocols~\cite{DechantEPL2017} to become a reality. Generally, substituting the thermal environment by engineered (quantum) reservoirs~\cite{MillenNJP2016} -- e.g., via cavity optomechanical methods~\cite{WollmannScience2015,PirkkalainenPRL2015,LecocqPRX2015} -- opens the door for investigating extended schemes with nonthermal or nonclassical baths, making it possible to bypass the Carnot efficiency limit~\cite{RossnagelPRL2014,KlaersPRX2017}. More importantly, these engineered reservoirs will enable researchers to understand and minimize entropy production in the finite time domain, which is paramount, in engines and elsewhere, to optimize efficiency and minimize wasted energy. In the quantum domain, coherence, entanglement, and measurement backaction, which are now starting to become observable in optical levitodynamics, will play key roles in engine design and performance, addressing open questions in quantum thermodynamics~\cite{TalknerPRE2016,GieselerEntropy2018}. By helping researchers to understand these concepts and providing flexible possibilities for spatiotemporal design, levitodynamics may inspire an entirely new generation of classical and quantum nanoscale mechanical engines.

\textit{Material science.}
A defining advantage of levitated particles is their high internal complexity as a mesoscopic solid-state system. The ability to investigate and control this isolated solid will open new avenues for materials science. Specifically, the absence of surrounding gas and substrate will enable precise measurements of physi- and chemisorption rates in unusually clean environments~\cite{RicciArxiv2021}. Moreover, the enhanced control provided by levitation can be exploited for the growth of novel crystalline heterostructures~\cite{KanePRB2010}. In the field of droplet levitation~\cite{KriegerChemSocRev2012}, the control techniques developed in levitodynamics will make it possible to scale down current experiments with microdroplets to the nanoscale. The preparation of nanodroplets with different chemical composition, viscosity, refractive index, and surface tension will facilitate the study of processes such as phase transitions, coagulation, and photochemical reactions, among others, as a function of external stimuli. In the future, levitodynamics could provide a nanolab for engineering the material properties of levitated nanoobjects.

\textit{Condensed matter physics.}
Levitodynamics also enables researchers to study previously unexplored regimes of condensed matter where solid-state excitations (e.g., phonons, electrons, magnons) exist in extreme isolation and confinement~\cite{RubioLopezPRB2018}. Acoustic phonons have received the most attention, owing to their ubiquity and highest-recorded quality factor. This provides acoustic modes with extremely long coherence times but also makes them difficult to probe experimentally. Future detection will be possible by means of Brillouin scattering or, in rapidly rotating spheres, by the change in polarizability stemming from centrifugal elastic deformation~\cite{HummerPRB2018}. In optical traps, this deformation would additionally give rise to a complex dynamical coupling between rotation, motion, and acoustic field, which could shed light on the motional decoherence induced by internal vibrations. Once detection schemes become available, acoustic phonons will represent a powerful resource. According to theoretical works, a tunable and strong coupling between acoustic phonons and the particle motion can be engineered by using mediating degrees of freedom (e.g., magnons in levitated magnets)~\cite{GonzalezBallesteroPRB2020}. This could enable the use of internal acoustic resonances for motional cooling without the need for external feedback, or for engineering internal and external states via motional control~\cite{GonzalezBallesteroPRL2020}, such as squeezed magnonic, acoustic, or motional states. Because the novel properties of acoustic phonons stem entirely from confinement and isolation, similar exotic properties can be expected for any other condensed matter excitations (e.g., semiconductor excitons in levitated quantum dots or supercurrents in levitated superconducting particles)~\cite{RomeroIsartPRL2012}. Despite the potential of levitated particles, both for fundamental solid-state science and as tools to control external dynamics, their internal physics remains largely unexplored.

\section{Open challenges}

The prospects outlined above have established levitodynamics as a vibrant research field. However, several challenges must be addressed before such research directions can be fully explored.

\textit{On-chip levitodynamics.} 
The implementation of levitation-based sensing platforms faces three substantial obstacles. The first one concerns the development of chip-scale levitation systems operating in high isolation. On-chip integration is key to interface levitodynamics with other existing technologies, to increase platform robustness, and to devise autonomous and portable sensors. Despite achievements in this direction, the transition from current bulky proof-of-principle experiments to compact on-chip platforms remains challenging. On the technical side, it requires the engineering of optical, magnetic, and electric fields, or combinations of them, at the chip surface to create microscale traps. These traps must include on-chip feedback-control schemes to stabilize traps in ultrahigh vacuum and/or reduce the ring-down time of trapped high-quality factor oscillators. Developing an autonomous integrated platform will also require miniaturization of key surrounding components, from vacuum equipment to cryogenic setup. On the fundamental side, particles in microscale traps experience surface-induced forces and decoherence, which are relevant at short particle-surface separations. These short-range interactions affect the trapping potential, and the dynamics they induce depend on the internal degrees of freedom of the trapped object and their properties (e.g., temperature, internal equilibration times). Because these degrees of freedom can be largely out of equilibrium, the forces may differ from typical predictions on the basis of quasi-equilibrium assumptions~\cite{RubioLopezPRB2018}. Second, in analogy to anomalous heating in trapped ions~\cite{BrownuttRMP2015}, surface-induced dissipative forces create additional heating. These forces can have multiple origins, such as Johnson noise or patch potentials in Paul traps~\cite{BrownuttRMP2015,TellerPRL2021} or eddy currents or hysteresis losses for magnets trapped above superconductors~\cite{PratCampsPRAppl2017,VinantePhysRevAppl2020,GieselerPRL2020}. In most cases, these mechanisms are not well understood for levitated particles. To implement on-chip levitation in high vacuum, future research must focus on developing a fundamental understanding of surface-induced forces and implementing strategies to mitigate them.

\begin{figure*}[t!]
	\centering
	\includegraphics[width=0.66\linewidth]{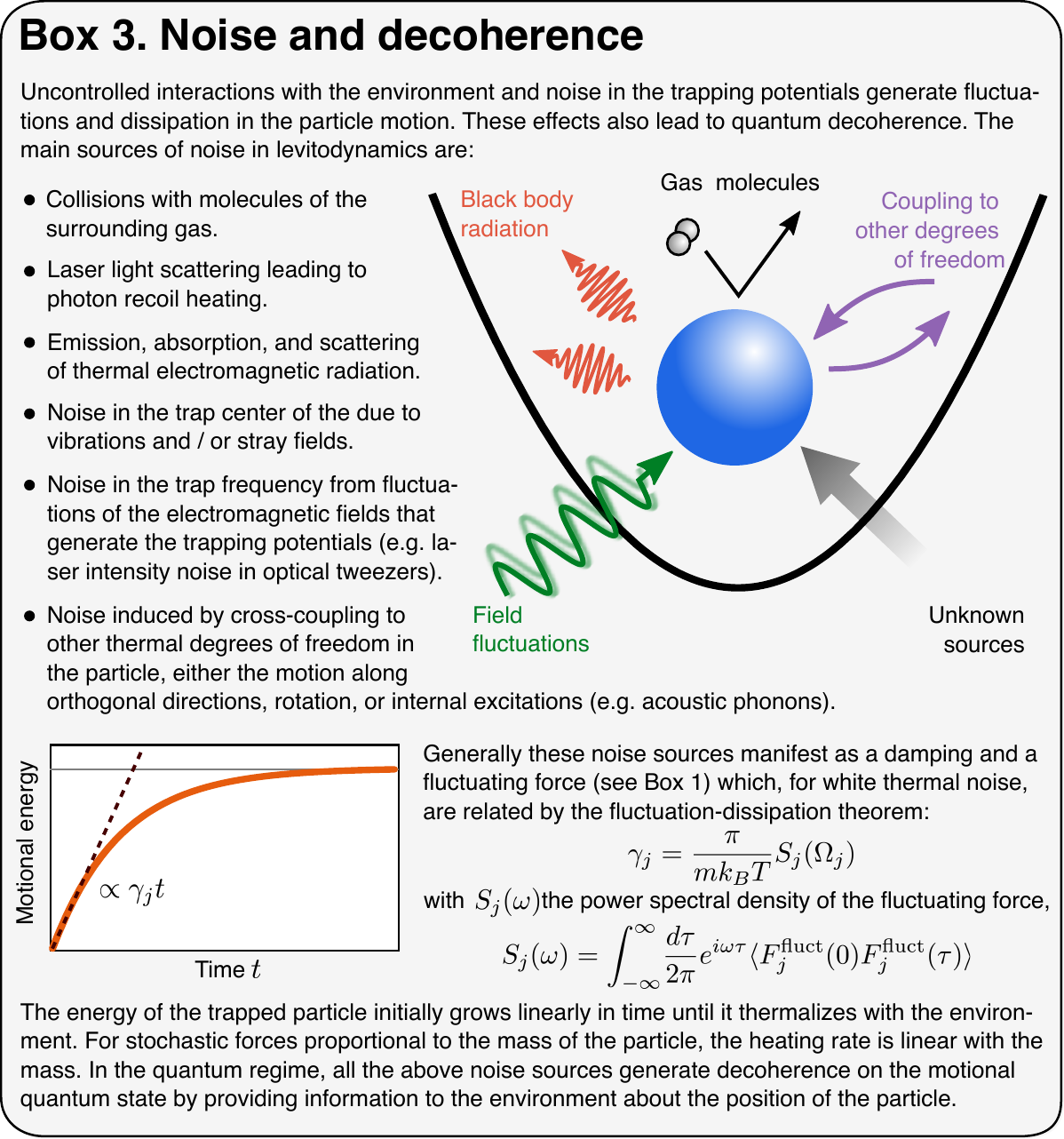}
	\vspace{-0.2cm}
	%\vspace{-0.2cm}
 	%\caption{\textcolor{purple}{CGB for myself: justify text, change particle color to blue, label ``Box 1: Levitation...''}}
\label{Box3Noise}
\end{figure*}

\textit{Sensor performance.}
The second challenge ahead for levitation-based sensors is to establish their true sensing potential. Most studies are at the proof-of-concept level and focus on the minimum detectable signal, commonly referred to as sensitivity. However, for industrial and field applications, other attributes -- such as robustness, reliability, and stability -- are more important. The stability of an instrument is characterized by the Allan minimum, corresponding to the signal integration time that yields the lowest variance~\cite{HebestreitRSI2018}. Undesired effects, such as laser pointing noise~\cite{MonteiroPRA2017}, eddy currents in magnetic levitation~\cite{TimberlakeAPL2019}, or technical noise~\cite{MonteiroPRA2020}, limit the stability of levitated sensors. Stability against long-term drifts is especially critical for low-frequency signal detection using high-quality factor levitated sensors. Technical noise is a particular obstacle for torque sensing with fast nanorotors because it linearly increases with rotational frequency and makes shot-noise-limited operation difficult~\cite{VanDerLaanPRA2020}. Other challenges include calibration uncertainties (e.g., particle mass)~\cite{HebestreitRSI2018,RicciNanoLett2019,BlakemorePRAppl2019,ZhengPRL2020} and the temperature dependence of system parameters (e.g., index of refraction of particle). Although some of these noise sources can be suppressed by differential measurements, the potential of levitation-based sensors for commercial applications requires thorough evaluation and optimization procedures.

\textit{Enhancing sensitivity.} The third important challenge pertains to boosting the sensitivity of levitation-based sensors, especially for uses in fundamental science (e.g., detection of weak forces or of quantum features). A fundamental sensitivity limit is reached when all thermal and technical noise sources are eliminated and the measurement becomes shot-noise limited. State estimation based on Kalman filtering~\cite{WieczorekPRL2015,MagriniNature2021} in combination with quantum control can be used to compensate for drifts and stabilize operation. Stretching the measurement precision beyond the standard quantum limit will require implementation of quantum-enhanced sensing protocols (e.g., squeezed states of light and/or motion). In the future, completely new sensing strategies will open up once it is possible to harness the “wave properties” of levitated particles and to capitalize on progress achieved in the field of matter-wave interferometry.

\textit{Decoherence.} Levitated objects are macroscopic and thus very sensitive to quantum decoherence (Box~\hyperref[Box3Noise]{3}). To perform quantum experiments, strategies to mitigate or circumvent such decoherence must be developed. Decoherence due to collisions with molecules of the surrounding gas can be suppressed by operating in ultrahigh vacuum, such that the probability to scatter a single gas molecule in each experimental run becomes negligible. The use of laser light also generates motional decoherence in two ways: by direct scattering of photons and by raising the internal energy of the particle through absorption, thereby increasing the emission rate and frequency of the thermal electromagnetic field radiated by the particle. In both cases, the photons (scattered and thermally emitted, respectively) carry away information about the center-of-mass position and orientation of the particle, thus generating decoherence. Controlling levitated particles without light does not necessarily reduce decoherence, as it requires either highly charged particles, which are more sensitive to stray electric fields, or magnetic or superconducting particles levitated using magnetic fields, which typically need on-chip schemes that are prone to surface-induced decoherence. The decoherence challenge is amplified as soon as the extension of the particle wave function grows, a required step toward exploration of macroscopic quantum physics. In addition, adding nonlinearities to the quantum motional control toolbox of a levitated particle, as required for exploration of non-Gaussian physics, typically involves either strong coupling to a two-level quantum system or the use of strongly nonharmonic potentials. In both cases, a levitated particle must be placed in a region of large field gradients, which unavoidably require the proximity of the particle to other materials and are hence accompanied by surface-induced decoherence. To tackle these challenges, quantum experiments with levitated particles, and particularly those exploring macroscopic quantum physics, will have to involve fast experimental runs and minimal use of laser light while avoiding large surfaces and time-dependent electromagnetic fields. A deeper understanding of decoherence in levitodynamics, attainable by the progress of experiments and theory, will most likely lead to innovative ideas for controlling levitated particles in the quantum regime.

\textit{Repeatability.} In addition to minimizing decoherence, quantum experiments must be robust and repeatable. A quantum tomography protocol involves (i) preparation of the initial state, (ii) state evolution, and (iii) measurement of the final state. This sequence must be repeated many times, with the same particle and under identical conditions, to extract the underlying quantum statistics. Because protocols likely involve sequences of free fall and changes in trapping potential, it is difficult to retrap the particle and bring it back to its initial state. Optimal control protocols, methods for low-noise measurements and filtering, and ultrafast data acquisition and processing techniques will thus be needed.

\textit{Scalability.} Scaling up the size and number of levitated particles (to enhance inertial sensitivities or to study many-particle physics, respectively) also poses several challenges. The more massive the particle, the more difficult it is to levitate in a high-frequency trap, the stronger its coupling to the environment, and the smaller its zero-point motion. This introduces additional noise and hinders motional ground-state cooling and all related applications (e.g., metrology of impulses or transient signals, tests of quantum mechanics)~\cite{MonteiroPRA2020}. Levitating several particles is also challenging, owing to the difficulty of simultaneously multiplexing different traps~\cite{CurtisOptComm2002} and controlling the resulting interactions between the particles contained in them (e.g., coupling that stems from optical binding~\cite{AritaOptical2018}). Once these complex traps are engineered, practical questions, such as those pertaining to controlled loading of several particles and the cooling required for applications, will have to be addressed. A step ahead in terms of complexity are hybrid multiplexed traps, which would enable levitation of particles of different sizes and types (e.g., dielectric, charged, magnetic, superconducting) for exploring, e.g., multispecies statistical mechanics, including complex (magnetic, electrostatic) interactions, or reconfigurable levitated particle lattices to model solid-state physics. Developing such hybrid multiplexed traps will likely require a combined effort toward hybrid trap design and on-chip integration.

% \vspace{-0.2cm}

\section{Outlook} 

% \vspace{-0.2cm}
Levitodynamics brings together appealing features from different fields -- examples include levitation techniques from atomic physics and biosciences, control techniques from atomic and molecular optics, the richness of solid-state physics, and the potential to couple motion to various degrees of freedom and/or external systems. It is the combination of all of these properties in a single platform that brings forth unique opportunities for science and technology. The path toward attractive research directions, from commercial sensing devices to fundamental quantum physics, has been unlocked by the many recent experimental achievements. Exploring the multiple future prospects for science and technology is a multidisciplinary endeavor, which will establish levitated objects in high vacuum as highly controllable nanolabs for use in areas as diverse as nonequilibrium physics, condensed matter physics, particle physics, and the foundations of quantum mechanics. 

\textcolor{white}{\colorlinks{white}{\cite{RusconiPRB2016,RusconiPRL2017,RusconiPRB2017}}}

\section*{Acknowledgments}

The authors acknowledge the valuable feedback from colleagues all over the world and thank C. Dellago in particular for suggesting the term “levitodynamics.” 

\textbf{Funding:} This work has received funding from the Q-Xtreme project of the European Research Council under the European Union’s Horizon 2020 research and innovation program (grant agreement 951234). \textbf{Competing interests:} None declared.
\newpage

\bibliography{scibib}
\end{document}